\documentclass[twocolumn,showpacs,preprintnumbers, reprint, amsmath, amssymb,prl]{revtex4-1}

\usepackage{graphicx}
\usepackage{dcolumn}
\usepackage{bm}

\graphicspath{{./}}

\begin{document}
\title{Entangled granular media}
\author{Nick Gravish}
\affiliation{School of Physics, Georgia Institute of Technology, Atlanta, Georgia 30332, USA}
\author{Scott V. Franklin}
\affiliation{School of Physics, Rochester Institute of Technology, Rochester, New York 14623, USA}
\author{David L. Hu}
\affiliation{Mechanical Engineering, Georgia Institute of Technology, Atlanta, Georgia 30332, USA}
\author{Daniel I.\ Goldman*}
\affiliation{School of Physics, Georgia Institute of Technology, Atlanta, Georgia 30332, USA}
\date{\today}

\begin{abstract} 
    We study the geometrically induced cohesion of ensembles of granular ``u-particles'' which mechanically entangle through particle interpenetration. We vary the length-to-width ratio $l/w$ of the u-particles and form them into free-standing vertical columns. In laboratory experiment we monitor the response of the columns to sinusoidal vibration (frequency $f$, peak acceleration $\Gamma$). Column collapse occurs in a characteristic time, $\tau$, which follows the relation $\tau = f^{-1} \exp(\Delta / \Gamma)$. $\Delta$ resembles an activation energy and is maximal at intermediate $l/w$. Simulation reveals that optimal strength results from competition between packing and entanglement.
\end{abstract}

\maketitle

Many living and nonliving materials are composed of ensembles of non-spherical particles. In the biological world, ant rafts \cite{mlot2011fire, *anderson2002self} and eagles nests \cite{hansell2000bird} are held together through the interpenetration of elongated and bent limbs and branches, respectively. In soft condensed matter systems composed of ensembles of particles, particle shape influences rheological and structural properties like viscosity \cite{BrownSuspension, *hsueh2010effective, *egres2005rheology}, yield stress \cite{BrownSuspension, kramb2011nonlinear} and packing density \cite{man2005experiments, blouwolff2006coordination}.

In studies of granular materials (GM), collections of athermal particles that interact through dissipative repulsive contact forces, most attention has been devoted to approximately spherical (convex) particle shapes \cite{*desmond2006jamming, *trepanier2010column, donev2004improving, philipse1996random}. Dry non-convex granular particles also interact through repulsive contact forces. However non-convex particles possess the ability to interpenetrate, and thus depending on shape may experience an effective cohesion through interpenetration. We refer to such interpenetrating ensembles as ``entangled'' ensembles. While the static packing of granular non-convex particles \cite{donev2004improving, *jiao2009optimal} and the rheology of non-convex particle suspensions \cite{BrownSuspension} have been previously studied, the role of particle entanglement in granular dynamics and stability is unexplored.

In this paper we report laboratory experiment,  computer simulation, and theory to elucidate the role of particle shape in the packing and stability of a model non-convex granular system of u-shaped particles (``u-particles''). Non-convex particles in the shape of a ``u'' are one of the simplest shapes in which concavity can be systematically controlled by varying arm length. Similar to previous studies \cite{jaeger1989relaxation,philippe2002compaction,herminghaus2005dynamics} \cite{clement2008spreading,*richard2005slow} of dry GM we use vertical vibration to probe relaxation and rearrangement of the u-particle ensembles, thereby characterizing the effective cohesiveness of the entangled ensembles. We show that non-convex particle shape leads to effects which are qualitatively different from concave dry GM and display phenomena similar to those observed in studies of wet cohesive granular materials \cite{herminghaus2005dynamics}.

\begin{figure}[b]
\begin{centering}
\includegraphics[scale=1]{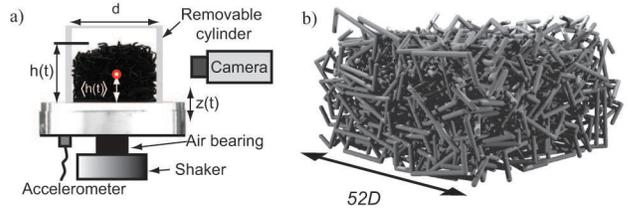}
\caption{Experiment and simulation of u-particle ensembles. a) Free-standing vertical columns are formed on a vibration table and the collapse dynamics are observed under vertical vibration. b) Computer generated ensembles of u-particles.}
\end{centering}
\end{figure}

{\em Methods}-- The u-particles in our experiments consisted of steel staples (Duo-fast; Vernon Hills, Il.) of constant width, $w$=1.17~cm, and variable length, $l$ ($l/w \in [0.02, 1.125]$), see Figs. 1 and 2. The cross section of all particles was rectangular with thickness of 0.5~mm and width 1.27~mm which corresponded to a rod-like aspect ratio for $l/w = 0.02$ particles of $\approx 14$. Particles with $l/w =0.02\pm 0.02,0.13 \pm 0.02, 0.15\pm 0.03 $ and $ 0.28\pm 0.04$ were cut to length by hand and the remaining sizes were purchased. Columns occupied a volume $V = \pi h (d/2)^2$, and the bulk volume fraction was calculated as $\phi = \frac{M}{\rho_{st} V}$ where $M$ is total particle mass and $\rho_{st} = 7.85~g/$cm$^3$ is the density of steel.

Sinusoidal vibration was generated by an electromagnetic shaker (VTS; Aurora, OH; Fig. 1a). The shaker was controlled by LabVIEW and a Tecron 7550 power amplifier. Acceleration of the vibration table was measured by an accelerometer (PCB Piezotronics; Depew, New York). Columns were subjected to vibration for a maximum of 20 minutes or until complete column collapse occurred.

Column collapse was monitored using a digital camera (Point-Grey; Richmond, BC, Canada). Image capture was triggered at constant phase in oscillation. Images were analyzed in Matlab.

In Monte-Carlo simulation (Fig. 1b) we studied the packing of ensembles of u-particles. U-particles were created from three sphero-cylinders, each of diameter $D$, placed at right angles with backbone width $W=14D$, measured from center to center of arms, and variable barb length $l/w \in [0, 1.3]$. We estimate the particle volume to be $v_p = \pi (D/2)^2 ( 2L + w +\frac{4}{3}(D/2)^2 )$ The cross-sectional particle shape differed between simulation and experiment (rectangular in experiment, circular with diameter $D$ in simulation). Excluded volume, $v_e$, was numerically estimated by determining the probability $p$ of overlap between two randomly oriented and positioned particles within a fixed spherical volume $V$. Overlap was determined by computing the minimum distance between all line segments composing the two particles; if this distance was less than the sphero-cylinder diameter, $D$, particles overlapped. Excluded volume is the product of overlap probability and sphere volume $v_e = pV$.

We used a packing algorithm to generate packings of non-overlapping u-particles. Packing proceeded in two steps: first particles were placed at random position and orientation inside a cubic volume of cross sectional area ($52D \times 52D$). If particle placement resulted in overlap the placement was rejected and a new position was randomly selected. If after 10,000 iterations a non-overlapping particle location was not found the algorithm proceeded to step two. In the second step randomly selected particles were displaced downwards a small random direction and distance $\frac{D}{10}$. If the new location of the particle resulted in particle overlap the particle was returned to the original location and a new particle chosen. The algorithm was halted after the center of mass height of the ensemble reached a steady-state. The volume fraction of the simulated packing was determined by measuring the average height of the pile to obtain the occupied volume and then dividing this by the total volume of particles.

{\em Column formation}-- In experiment, collections of monodisperse particles with fixed $l/w$ were formed into free standing cylindrical columns with column diameter, $d$=4.4~cm, and height, $h_0$ = 3~cm. Columns were prepared by pouring particles into a cylindrical container followed by a 20 s sinusoidal vibration of the base at a frequency, $f = 30~Hz$, and peak acceleration, $\Gamma = 2$ (in units of gravitational acceleration $g$). We confirmed that steady state volume fraction was reached through our preparation protocol in separate experiments conducted over a 60 s time period.

The final volume fraction of the column following packing, $\phi_f$, was a monotonically decreasing function of $l/w$ (Fig. 2b). The value of $\phi_f$ observed for $l/w=0.02$ particles was within the range observed in cylindrical rod experiments with comparable aspect ratio (length/thickness $\approx 14$) which pack to $\phi_f = 0.28-0.34$ depending upon preparation \cite{wouterse2009contact, philipse1996random, blouwolff2006coordination,desmond2006jamming}.

Following the packing preparation we removed the confining container leaving the column free-standing (Fig. 3a). During removal of the confining cylinder the $l/w=0.02$ particles were marginally stable with partial column collapse occurring approximately $50\%$ of the time, similar to the results reported in \cite{trepanier2010column}. Spontaneous collapse of the $l/w>0.02$ columns was rarely observed.

\begin{figure}[t]
\begin{centering}
\includegraphics[scale=1]{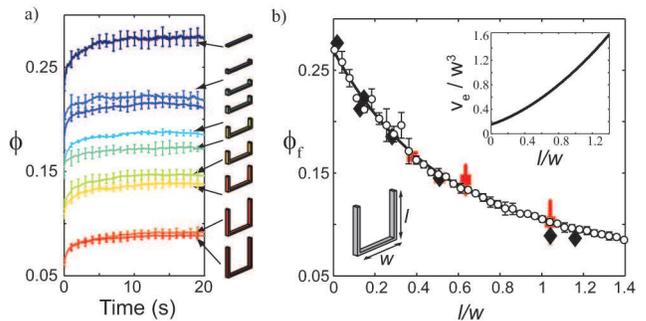}
\caption{Packing of u-particles. a) $\phi(t)$ during column preparation. b) Final packing fraction, $\phi_f$, as a function of particle geometry in experiment (column diameter d = 4.4 cm diamonds and d = 5.6 cm squares) and simulation (white circles). Line is theory prediction from the random contact model. (Inset) Excluded volume of sphero-cylinder u-particles from simulation.}
\end{centering}
\end{figure}

{\em Column collapse}-- We applied sinusoidal vibration to the base of the free-standing column and observed the collapse process. We characterized collapse dynamics by monitoring the centroid height, $h(t)$, of the column (Fig. 3). The collapse duration decreased with increasing $\Gamma$ (Fig. 3b) and collapse dynamics were well described by a phenomenological stretched exponential fit function $\frac{h(t)}{h_0} = e^{[-(\frac{t}{\tau})^{\beta}]}$. The parameter $\tau$ is the characteristic collapse time and $\beta$ is the stretching parameter \cite{phillips1996stretched}. The stretched exponential function is frequently applied to the description of relaxation dynamics of disordered systems \cite{phillips1996stretched}.

\begin{figure}[b]
\begin{centering}
\includegraphics[scale=1]{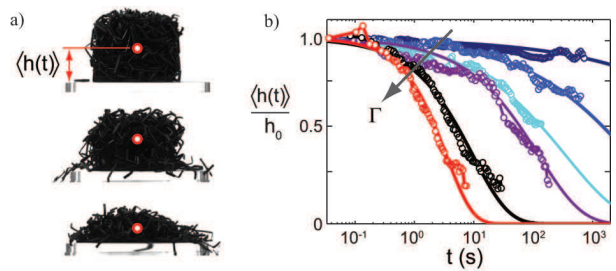}
\caption{Collapse dynamics. a) The normalized centroid height, $h(t)/h_0$ (red dot) of the column during collapse. Vibration parameters are $\Gamma=2$, f = 30Hz, and images are separated by 90 oscillation periods. b) Relaxation of $h(t)/h_0$ is shown for $l/w = 0.379$ for $\Gamma = 1.23, 1.48, 1.70, 1.96, 2.20, 2.53$ respectively. Fit lines are stretched exponentials given in the text.}
\end{centering}
\end{figure}

Consistent with previous studies \cite{mattsson2009soft, philippe2002compaction} $\beta$ was in the range of 0.5 - 1 and decreased slightly as $\Gamma$ increased but was independent of particle geometry. For fixed $l/w$, $\tau$ decreased with increasing $\Gamma$. The logarithm of $\tau$ increased with $1/\Gamma$ (Fig. 4a) and was fit by an exponential $\tau = f^{-1}e^{\Delta/\Gamma}$ with $\Delta$ as the single fit parameter.

The exponential fit is indicative of an Arrhenius-like process observed in the relaxation of activated systems. The Arrheniuse process describes the escape probability of a thermally or mechanically activated particle from a potential well of depth $\Delta$. In thermal systems the escape time is proportional to one over the Boltzmann factor $\exp(-\frac{E}{kT})$ where $E$ is the activation energy required to overcome the potential barrier. In our system thermal effects are negligible, and we hypothesize that the mechanical excitation plays the role of a thermal energy source ($\Gamma$ analogous to $kT$) and $\Delta$ is analogous to an energy barrier resulting from particle entanglement.

\begin{figure}[t]
\begin{centering}
\includegraphics[scale=1]{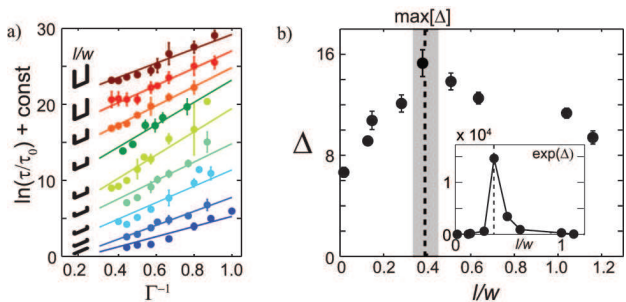}
\caption{Particle shape and optimal geometric cohesion. a) The logarithm of the relaxation time versus inverse acceleration with exponential fit lines $\tau = f^{-1}e^{\Delta/\Gamma}$  ($\tau_0 = 1$~s). Curves are offset vertically for clarity. Error bars are standard deviation of 4 or greater replicates. b) $\Delta$ as a function of $l/w$. Dashed line indicates estimated maximum of $\Delta$ (see [16]). Error bars represent 95\% confidence interval of the best fit lines from (a).}
\end{centering}
\end{figure}

Column collapse occurred through the separation of entangled particles during vibration. We therefore expected that the hindrance of motion due to particle entanglement---and thus $\Delta$---would increase monotonically with the size of the concave region and thus particle length. Instead we found that $\Delta$ was a non-monotonic function of $l/w$ (Fig. 4b) with $\Delta$ reaching a maximum value at intermediate $l/w=0.394 \pm 0.045$ \footnote{We estimate the maximum and standard deviation of $l/w$ in experiment using a weighted average of points near the peak}. $\Delta$ appears in an exponential and thus the relaxation time for fixed $\Gamma$ displays a surprising sensitivity to variation of particle shape (Inset Fig. 4b). We posit that the maximum in $\Delta$ is related to the statistics of particle entanglement within the bulk, and we next study entanglement propensity in theory and simulation.

\begin{figure}[t]
\begin{centering}
\includegraphics[scale=1]{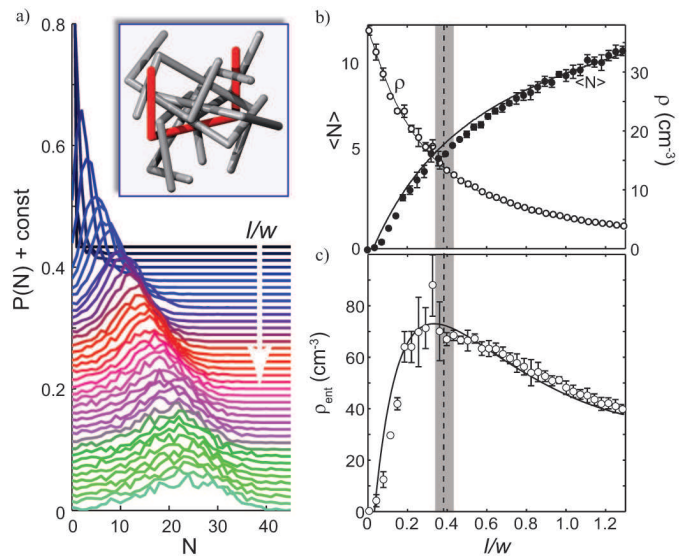}
\caption{Statistics of particle entanglement in simulation. a) The probability distribution of entanglement number, $N$, as a function of $l/w$. Curves are shifted vertically for clarity; $l/w = 0$ at top and increases in increments of 0.036 down. Inset: red particle containing 6 entangled neighbors. b) Mean values for $N$ and $\rho$ measured in simulation (circles) and the theoretical fit (black line). c) Density of entanglements as a function of $l/w$ and the theoretical fit (black line). Vertical dashed line and gray bar correspond to the mean and standard deviation of the estimated maximum of $\Delta$ from experiment.}
\end{centering}
\end{figure}


{\em Model and simulation}-- In simulation we generated u-particle ensembles of varying $l/w$ (Fig. 1b) and measured the packing fraction, $\phi_f$. We found excellent agreement between experiment and simulation in $\phi_f$ (Fig. 2b). The scaling of $\phi_f$ with $l/w$ can be understood through the random contact model originally developed to describe the packing of hard rods \cite{philipse1996random}.

Assuming a homogeneous packing of particles, the random contact model relates the excluded volume, $v_{e}$, occupied volume, $v_p$, and average number of particle contacts, $C$, to the bulk packing fraction $\phi_f$ through $\phi_f = C\frac{v_p}{v_{e}}$.

We numerically computed $v_{e}$ (Inset Fig. 2b) and found excellent agreement between the prediction of the random contact model and the values of $\phi_f$ measured in experiment and simulation (Fig. 2b). Fitting the random contact model to experimental and simulation packing data we obtained $C = 8.74 \pm 0.04$, close to the value found in studies of colloid and granular rod packing \cite{philipse1996random, blouwolff2006coordination, wouterse2009contact, desmond2006jamming}.


We hypothesized that particle entanglement within the column would influence the relaxation time during vertical vibration. Thus we expected that the maximum in $\Delta$ should correspond to a maximum in the density of particle entanglements. In simulation we defined two particles as entangled when the center line of one particle intersected the internal plane of the neighboring particle (Inset Fig. 5a). We measured the number of entanglements per particle, $N$, for each particle in simulation. The probability distribution function, $P(N)$, was sensitive to $l/w$ (Fig. 5a) with mean value $ \langle N \rangle$ increasing monotonically with $l/w$ (Fig. 5b). The increase was sub-linear indicating that $ \langle N \rangle$ grew slower than that of the particle's convex area $(l-D)(w-2D)$.

The scaling of $\langle N \rangle$ with $l/w$ can be determined by considering the solid volume occupied by the entangled particles in the focal particles convex region (the convex area with infinitesimal thickness $\delta$). Assuming a homogeneous packing the solid volume in this region is $V_{ent} = \phi_f (l-D)(w-2D) \: \delta$. Since each entangled particle contributes only a portion to $V_{ent}$ in the shape of an ellipse of thickness $\delta$, on average $V_{ent} = \alpha \langle N \rangle \pi \delta \frac{D^2}{4}$ where $\alpha > 1$ accounts for the non-planar crossings (Inset Fig. 5a). Solving the above relations yields $\langle N \rangle = \frac{4C}{\alpha} \left(\frac{v_p (l-D) (w-2D)}{\pi v_{e} D^2}\right)$. With single fit parameter, $\alpha = 2.648 \pm 0.108$, we find excellent agreement between the predicted number of entanglements per particle and those measured in simulation (Fig. 5b).

The spatial density of particle entanglements is $\rho_{ent} = \langle N \rangle\rho$ where $\rho = \frac{C}{v_{e}}$ is the number density of particles (Fig. 5b). Substitution for $\langle N \rangle$ yields $\rho_{ent} = \frac{4C^2}{\pi \alpha}\left(\frac{v_p (l-D) (w-2D)}{v_{e}^2 D^2}\right)$ and again the simulation and theory are in good agreement (Fig. 5c) using the previosly determined fit parameters $C$ and $\alpha$. Furthermore the experimental maximum max$[\Delta]$ at $l/w=0.394 \pm 0.045$ is close to the value obtained in simulation and theory of $l/w = 0.340 \pm 0.015$ suggesting that the large relaxation times for the intermediate u-particle columns is due to the large density of mechanical entanglements.

\textit{Conclusion}-- Similar to rod-like particles \cite{wouterse2009contact, philipse1996random, blouwolff2006coordination, desmond2006jamming}, columns formed from u-particles are stabilized through the inhibition of particle rotation and translation among the entangled particles. The addition of the transverse ends which form concave u-particles leads to mechanical entanglement and increases column stability. However the increase in entanglement with increasing length is offset by the decrease in particle packing density. These two trends conspire to generate a maximum in the density of mechanical entanglements in collections of non-convex particles of intermediate $l/w$--thus columns of these particles most strongly resist separation.

The random contact model utilized to explain the optimum geometry for entanglement of u-shaped particles assumes only uncorrelated particle contacts within the bulk. Thus we expect the results to apply to rigid non-convex particulate systems of all scales. For instance a recent study of suspension rheology found that convex particles of differing shape collapsed to a viscosity-stress master curve while concave particle did not collapse to this curve; this difference was attributed to particle entanglement effects \cite{BrownSuspension}. At the micro-scale, polymers with rigid pendants oriented perpendicular to the polymer chain increase internal molecular free volume and hinder polymer motion which significantly affects rheology similar to geometric entanglement \cite{tsui2006minimization}. At the macroscale strain-stiffening of model polymers is associated with entanglement \cite{BrownStiffening}. Even organisms can benefit from geometric entanglement. For example the fire ant \textit{Solenopsis invicta} and the army ant \textit{Eciton burchelli} create waterproof rafts and shelters--which have been described as akin to living chain mail \cite{mitchell2009complexity}--through the interlocking and entanglement of limbs and mandibles \cite{mlot2011fire, anderson2002self}.

{\em Acknowledgements}-- The authors would like to acknowledge Paul B. Umbanhowar for helpful discussion and Geoff Russell, Luis Saldana, and Nathan Jankovsky for help with experimental fabrication. Funding support for N.G. and D.I.G. provided by NSF Physics of Living Systems $\#$0957659. S.V.F. was supported by (NSF DMR-0706353). D.H. gratefully acknowledges the financial support of the NSF (PHY-0848894).

\end{document}